\documentclass{aa}
\usepackage{amsmath}
\usepackage{graphicx}
\usepackage{subcaption}
\usepackage{mdframed}
\usepackage{tensor}
\usepackage{mathtools}
\usepackage{natbib}
\usepackage[varg]{txfonts}
\bibpunct{(}{)}{;}{a}{}{,}

\DeclareMathOperator\erf{erf}
\usepackage{color}

\begin{document}

\title{Magnetic thread twisting in a simulated solar atmosphere}
\author{C. Sumner \inst{\ref{inst1}\ref{inst2}}\and Y. Taroyan\inst{\ref{inst1}}}
\institute{Department of Physics, University of Aberystwyth, Aberystwyth SY23 3BZ, Wales, UK \label{inst1}
\and
\email{CSumner@physics.org}\label{inst2}}
\date{Received date /
Accepted date }
\abstract {Plasma inflows accompany a variety of processes in the solar atmosphere such as heating  of coronal loops and formation of prominences. }
{We model a stratified solar atmosphere, within which a simulated prominence thread experiences density accumulation via a plasma inflow designed to mimic the formation process. We aim to investigate the interaction of such a system with torsional perturbations, and decipher the possible consequences.}
{We integrated the linearised equations of motion and induction in order to analyse the spatial and temporal evolution of torsional perturbations that are randomly driven at the photospheric footpoints.}
{Our results demonstrate that magnetic threads will experience twist amplification. Different sources and sinks of energy and the corresponding amplification mechanisms are identified. Threads reaching chromospheric heights are most susceptible to magnetic twisting with the maximum twist occurring near their footpoints.  The amplifying twists are associated with a standing wave behaviour along the simulated threads.}
{Our work suggests that torsional perturbations may be amplified within prominence threads, with strong magnetic twists forming at the footpoints. The amplification process is facilitated by small length scales in the background magnetic field. On the other hand, a small length scale in the background density inhibits growth. Possible consequences of the amplified twists  are
discussed, including their role in supporting the dense plasma within a prominence structure.}
\titlerunning{Magnetic thread twisting in a simulated solar environment}
\authorrunning{C. Sumner \& Y. Taroyan }
\maketitle 
\section{Introduction}\label{Introduction}
The solar atmosphere is a dynamic environment dominated by magnetic structures at all scales. Within the lower atmosphere, chromospheric fibrils are observed, with larger arch structures observed at higher altitudes. All of these structures are dependent upon magnetic fields for their shape, with plasma motions often travelling along them by magnetohydrodynamic (MHD) flows.
One such common feature are solar prominences, which are suspensions of cool and dense chromospheric plasma found in the corona. These prominences require a supporting magnetic topology, and are filled by associated plasma flows. The supporting magnetic fields form along polarity inversion lines in the presence of an over-arching magnetic arcade, and may either emerge from the photosphere or otherwise be formed by shearing motions \citep{Parenti2014}.

A filament thread may be filled via a variety of processes. In the event of an emerging supportive topology, an elevation of plasma may occur. Alternatively, they may be filled through dynamic heating or by direct injection of plasma along magnetic fields. Coronal magnetic threads and dips have also been modelled as filling by thermal non-equilibrium (e.g. \citet{Xia2016} and \citet{Luna_2012}), with the formation and draining of plasma blobs, as well as the formation of thin prominence threads that maintain the mass of the structure because of their dipped magnetic topology.

The filling of prominence threads may occur via chromospheric evaporation and capture. Plasma at the footpoints undergoing heating may evaporate and trigger upflows captured by magnetic flux tubes created along polarity inversion lines above the photosphere. Above the footpoints of these flux tubes, this evaporation will eventually result in density accumulation, which by a loss of thermal equilibrium can result in runaway cooling and condensation. The condensation associated with this process may be captured by the local topology of local flux tubes, and this results in the formation of chromospheric suspensions. The process of plasma accumulation along magnetic threads may occur through other mechanisms; for example, by shocked siphon flows because of their non-linear coupling with amplified magnetic twists \citep{Williams2018}.

Filament threads, which make up prominence structures, are subject to coherent large-scale oscillations across large and small scales. Such motions may be caused by, for example, nearby flaring activity on one scale, or by plasma motions within the thread on another. A review of prominence oscillations can be found in \cite{Arregui2018}. Torsional motions along a filament thread have been reported by \cite{Srivastava2017} using the Swedish Solar Telescope, and the existence of transverse waves in quiescent prominences was well documented by \cite{Hillier2013} using the Hinode Solar Optical Telescope. In the presence of such a dynamic system, Alfv\'enic waves are confirmed to be present, such as in the observations reported by \cite{Jess2009}.

Within the footpoints of prominences, tornado-like features have been observed by \cite{Su_2012} during formation of the wider structure, potentially contributing as an initial source of plasma \cite{Wedemeyer_2013}. These same structures were demonstrated to potentially represent an energy channel from the photosphere into the corona by \cite{Bohm2012} with torsional Alfvén waves propagating from these structures into the transitional region, with an associated transfer of energy \citep{2019NatCo..10.3504L}. An in-depth observational event of such a structure was presented by \cite{Morgan2012}, and by earlier studies such as that by \cite{Liggett1984} who were able to identify rotation in 5 of 51 studied prominences.

This paper serves to further the investigation from our previous work \citep{Sumner2020}, which demonstrated that a small-amplitude torsional Alfvenic pulse will be amplified in the presence of a plasma inflow along a magnetic thread. The previous investigation demonstrated, both analytically and numerically, that a flux tube thread with uniform density in a uniform magnetic field, which filled continuously along the length of the thread, would amplify any initial torsional perturbation. This amplification was the result of a wave-flow coupling between the plasma inflow and the Alfv\'en waves, and was therefore seen both in the azimuthal velocity and magnetic field within the plasma. This effect was demonstrated to be a fundamental feature of plasma physics that could occur in the absence of any external azimuthal magnetic or energy influx, with the prescribed inflow lacking any torsional motions across the boundary. This same effect along a similar thread was demonstrated by \cite{Taroyan_2019}, who used driven boundaries and demonstrated that the effect was not a result of resonance.
The current work expands on the earlier study. Here, we conduct a numerical investigation in the presence of simulated solar conditions. We aim to study how incorporating more realistic features of the solar atmosphere affects the amplification process and its consequences.

The paper is structured as follows: the model construction and the governing equations of the twists are presented in Section \ref{Model}. The numerical results are presented in Section \ref{Analysis}, and discussed in Section \ref{Discussion}. Finally our results are summarised in Section \ref{Conclusions}.

\section{Model}\label{Model}
We aim to simulate a prominence thread in the solar atmosphere, with footpoints rooted in the photosphere. The straightened thread model has length $L$, with photospheric boundaries at $\pm \frac{L}{2}$. We simulate an arch reaching into the solar atmosphere by varying the magnetic field strength and density along the thread in order to mimic the effects of gravitational stratification on background conditions. We additionally simulate mass loading and central accumulation of density as a result of a symmetric plasma inflow through the photospheric footpoints at $\pm \frac{L}{2}$ that has a stagnation point at $s=0$. Details of these are given below in this section. The analysis is carried out in the linear regime, and therefore the back-reaction effects on the plasma inflow through the non-linear ponderomotive force are not present.

We use the governing equations for axisymmetric motions derived by \cite{Hollweg1982}. We note that these equations can be reproduced by requiring that the r and z components of the magnetic field be time-independent \citep{Taroyan2021}. The magnetic field is therefore assumed to be strong enough to suppress any expansion or contraction of the magnetic field that would lead to temporal variations in the corresponding components \citep{Hollweg1992}.

Our 1.5D MHD model is used to investigate the evolution of torsional motions that are triggered either as a result of an initial perturbation or through footpoint drivers mimicking random footpoint buffeting. We investigate the spatial and temporal evolution of these torsional perturbations along the stratified thread. For small perturbations investigated in the linear regime, magnetic twists are decoupled from other perturbations. 

The plasma inflow has been shown previously to introduce a wave-coupling term which may amplify both initial torsional perturbations in the plasma, and those introduced through a driven footpoint, leading to an amplifying twist along the thread \citep{Taroyan_2019, Sumner2020}. This amplification was demonstrated analytically and numerically without the need for an influx of azimuthal magnetic field or energy \citep{Sumner2020}.  Previously, amplification was demonstrated in a system possessing a uniform background magnetic field and density, with a prescribed density accumulation which increased evenly along the thread. 

We extend the model in this work by introducing a background density and magnetic field which decay with height. The process of mass loading is no longer spatially uniform. In introducing this more representative solar background, the simulated torsional motions in the atmosphere will be subject to additional centrifugal and magnetic tension forces. In this work, we aim to investigate whether a plasma inflow can amplify torsional motions in conditions representative of the solar atmosphere, and where such amplification might be present in the linear regime.

\subsection{Time-dependent background}
As opposed to our previous work \citep{Sumner2020}, where the initial density and magnetic fields were constant along the thread, here we consider a stratified background where the density and magnetic field decay with height. The uniform density increase present in the previous work is also updated to more accurately represent prominence density accumulation.

We begin with the mass continuity equation:

\begin{equation}\label{Eq_Continuity}
\frac{\partial }{\partial t} \left( \frac{\rho}{B_s}\right) + \frac{\partial }{\partial s} \left( \frac{\rho }{B_s} U \right) =0,\\
\end{equation}
where $s$ is the curvilinear coordinate along the background magnetic field $B_s$, $\rho$ is density, and $U$ is the background flow.
The variation of the magnetic field is determined by the following solenoidal condition, as derived by \cite{Taroyan2021}:
\begin{equation}\label{solenoidal}
B_s r = \text{constant},
\end{equation}
with $r$ describing the distance of the thread from the central axis of symmetry. The solenoidal condition (\ref{solenoidal}) describes field lines that remain close to the axis of symmetry. We note that this condition is different from the one proposed by \cite{Hollweg1982} which may still be applied to a radially expanding field. This novel condition is consistent with the remaining governing equations.

The time-dependent plasma density within the thread is controlled by two length scales. The first, $\sigma_1$, determines the initial density contrast between the thread footpoint and apex, yielding a gravitationally stratified setup. The values of $\sigma_1$ used in our simulations were chosen to yield density contrasts between the chromosphere and corona consistent with one-dimensional hydrostatic models (\citet{Aschwanden2002}, \citet{Taroyan2006}, \citet{Price2015}) for describing the solar atmosphere. The density contrast between the photosphere and the chromosphere is about three orders of magnitude, consistent with VALiiiC \citep{Vernazza1981}.

\begin{figure}\label{img_model}
\centering
\resizebox{\hsize}{!}{\includegraphics{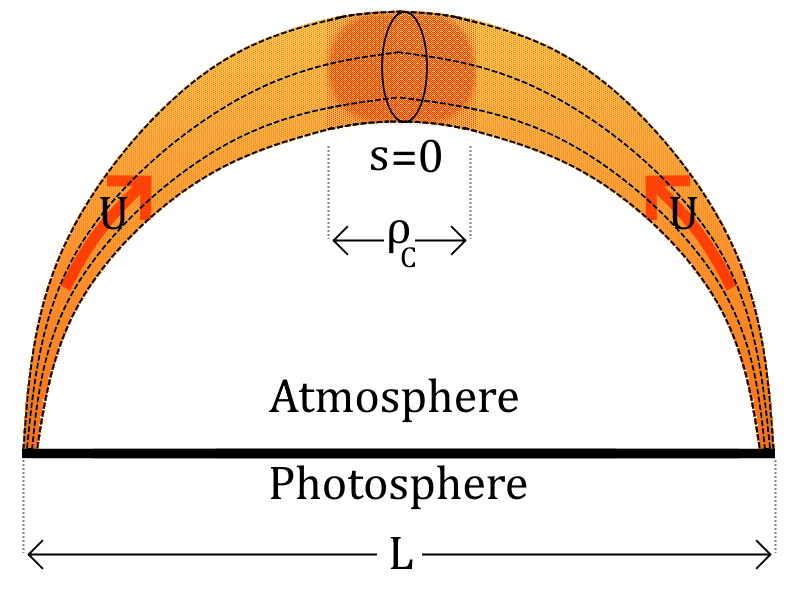}}
\caption{Stratified prominence thread as a single magnetic field line of length $L$, with boundaries at $s = \pm \frac{L}{2}$, anchored in a photosphere. A symmetric time-dependent plasma inflow, $U$, is introduced along the thread with a stagnation point at $s=0$ and a corresponding density profile with accumulation centred along the centre of the thread. Driven boundaries introduce torsional motions in the plasma through the footpoints.}
\label{fig:model}
\end{figure}

To this time-independent background, we then add a time-dependent component as a central accumulation, $\rho_c$, the width of which is determined by the length scale $\sigma_2$ (Figure \ref{fig:model}). By choosing large values of $\sigma_2$ we can approximate an increase across the whole thread. The parameters $\sigma_1$ and $\sigma_2$ are constants in each simulation. When combined, our density function is given as
\begin{equation}\label{Eq_Density}
\rho = \rho_b\frac{ \exp\left(\frac{s}{\sigma_1}\right)^2 + \frac{t}{t_0} \cdot \exp\left(-\frac{s}{\sigma_2}\right)^2}{\exp\left(\frac{L}{2 \sigma_1}\right)^2},
\end{equation}
such that
\begin{equation}\label{In_Density}
\rho_0 = \rho_b \exp\left(\left(\frac{s}{\sigma_1}\right)^2 - \left(\frac{L}{2 \sigma_1}\right)^2\right),
\end{equation}
which is the initial background density. In the follow-up simulations, density along the thread is normalised with respect to the initial footpoint density, $\rho_b$. The timescale, $t_0$, is used to control the rate of plasma accumulation.

The third length scale, $\sigma_3$, controls the gradient of the time-independent background magnetic field strength, $B_s$. Like the other $\sigma$ values, this value is fixed during the simulations, resulting in a time-independent profile for $B_s$.
\begin{equation}\label{Eq_Magnetic}
B_s = B_b \exp\left(\left(\frac{s}{\sigma_3}\right)^2 - \left(\frac{L}{2 \sigma_3}\right)^2\right).
\end{equation}

Similar to the density profile, the value at the footpoints, $B_b$, is used to normalise the magnetic field along the thread. Finally, speed is normalised with respect to the Alfv\'en speed, $v_{Ab}=\frac{B_b}{\mu_0 \rho_b}$, and time is normalised with respect to the Alfv\'en travel time $t_A=L/v_{Ab}$. 

The gravitational stratification of our simulated thread is determined by the length scales, $\sigma_1$ and $\sigma_3$. The time-dependent evolution of the background is determined by the length scale, $\sigma_2$, which controls the accumulation of plasma along the thread. The prescribed length scales aim to approximate the process of plasma accumulation in a stratified prominence thread.

The modelling of a three-dimensional changing topology of a flux tube in response to torsional motions is beyond the scope of this work. However, these changes will only appear in the non-linear stage of evolution, whereas the present work is carried out in a linear regime. The current model could still provide some clues as to the possible non-linear consequences, such as the subsequent plasma dynamics or the retention of the central density enhancement. By changing the density and magnetic field gradients along the field, we may simulate threads that reach different heights from the photosphere. This is important in terms of the expected thread dynamics. For example, \cite{Luna_2015}  argued that low-lying threads with dips may be those most capable of supporting density enhancement and contributing to prominence formation whereas those that reach coronal heights will experience temporary density enhancements in the form of transient plasma blobs. 

Finally, we construct a corresponding inflow velocity according to the continuity equation (\ref{Eq_Continuity}):
\begin{equation}\label{Eq_Inflow}
U = -\frac{\sigma_2 \sigma_3}{2 t_0} \sqrt{\frac{\pi}{\sigma_2^2 + \sigma_3^2}}
\frac{\erf\left(s \frac{\sqrt{\sigma_2^2 + \sigma_3^2}}{\sigma_2 \sigma_3}\right)
\exp\left(\frac{s}{\sigma_3}\right)^2}{\exp\left(\frac{s}{\sigma_1}\right)^2
+ \frac{t}{t_0} \cdot \exp\left(\frac{-s}{\sigma_2}\right)^2},
\end{equation}
where $\erf$ denotes the error function. This equation demonstrates that the flow is symmetric about $s=0$, because the error function is an odd function, and it decays with time. A cartoon representing this setup is shown in Figure \ref{fig:model}. Specific examples used in our numerical simulations are discussed further in Section \ref{Sec_BC}.

The physical motivation for the above choices of density and magnetic field is the corresponding contrast between the photosphere and the atmosphere caused by gravitational stratification. The mathematical motivation is the ease of controlling the spatial variation of the background magnetic field and the density through the length scales $\sigma_1$ and $\sigma_3$. The third length scale, $\sigma_2$, and the additional parameter, $t_0$, control the extent of plasma condensation and the rate of plasma filling, respectively. These controlling parameters are chosen such that they are consistent with values known from hydrostatic models or observations. The flow velocity in Equation \eqref{Eq_Inflow} is self-consistently determined by solving the continuity equation \eqref{Eq_Continuity} with the prescribed profiles for the density and the background magnetic field. 

A fully self-consistent approach would require modelling the formation of a prominence via the above-described processes of heating and evaporation to determine the profiles of density and flow velocity. This approach would mean solving the equation of mass continuity, the longitudinal component of the momentum equation, and the equation of energy with the inclusion of a heating term as well as thermal and radiative losses. In our simplified approach, we ignore the full set of equations as long as the amplitudes of the perturbations remain small.

\subsection{Torsional perturbations}
The evolution of the torsional motions in our simulation is governed by the corresponding MHD equations of motion and induction in a curvilinear coordinate system. These equations are complemented with the solenoidal condition (\ref{solenoidal}) that defines the expansion of the thread. As we focus on the linear phase of development, the governing equations are decoupled from the other MHD equations. Their derivation is presented in the appendices:
\begin{equation}\label{Eq_Induction}
\frac{\partial b_\theta}{\partial t} + 
\frac{\partial}{\partial s} \left( b_\theta U \right)=
\frac{\partial}{\partial s} \left( B_s v_\theta \right),
\end{equation}
\begin{equation}\label{Eq_Motion}
\frac{\partial v_\theta}{\partial t} + B_s U \frac{\partial}{\partial s} \left( \frac{ v_\theta}{B_s} \right) = \frac{B_s^2}{\mu_0 \rho}\frac{\partial}{\partial s} \left( \frac{b_\theta}{B_s} \right),
\end{equation} 
where the subscript $\theta$ represents the azimuthal coordinate, and $\mu_0$ is the magnetic permeability in a vacuum. Further, $b_\theta$ and $v_\theta$ represent twists in the magnetic field and the plasma associated motions, respectively. The remaining terms describe the time-dependent background introduced in the previous subsection. 
Here, we consider the linear evolution of torsional perturbations in a time-dependent background that are fully described by the induction and momentum equations (\ref{Eq_Induction}) and (\ref{Eq_Motion}). We adopt a linear approach where the torsional perturbations are decoupled from the longitudinal perturbations. However, additional equations should be included in any non-linear analysis, where back-reactions arising from gas pressure and the ponderomotive force are likely to further impact the underlying background conditions.

From these governing equations, the following wave energy equation may be derived:
\begin{equation}\label{InflowBCEnergy} 
\frac{\partial}{\partial t}\left(\frac{W_T}{B_s}\right) + \frac{\partial }{\partial s} \left( \frac{F_W}{B_s} \right) = \frac{U}{B_s}\frac{\partial \ln r}{\partial s} \left(\frac{b_\theta ^2}{\mu_0} - \rho v_\theta ^2 \right) -\frac{\partial}{\partial s}\left(\frac{U}{B_s}\right) W_m,
\end{equation}
where $W_T$ is the sum of the kinetic and magnetic energy densities:
\begin{equation}\label{InflowBCWT}
W_T = W_k + W_m = \frac{\rho v_\theta^2}{2} + \frac{b^2_\theta}{2 \mu_0},
\end{equation}
and $F_W$ represents the corresponding energy flux: 
\begin{equation}\label{InflowBCFW}
F_W = \frac{U}{B_s} W_T - \frac{v_\theta b_\theta}{2 \mu_0}.
\end{equation}
Integrating equation (\ref{InflowBCEnergy}) and using the divergence-free condition (\ref{solenoidal}), we obtain: 
\begin{eqnarray}
\nonumber
\frac{\partial }{\partial t}\int W_T r ds = -\int \frac{\partial }{\partial s}\left( F_W r\right) ds - \\
\label{integrated_energy}
\int 2 U r \frac{\partial \ln r}{\partial s}\left( W_m - W_k \right) ds - \int W_m \frac{\partial }{\partial s}\left( U r \right) ds, 
\end{eqnarray}
where the integration is carried out between the footpoint boundaries at $s=\pm\frac{L}{2}$. Equation (\ref{integrated_energy}) determines the temporal variation of the total azimuthal energy. The first term on the right-hand side of equation (\ref{integrated_energy}) represents the net azimuthal energy influx. The sources on the right-hand side of equation (\ref{integrated_energy}) represent the sum of the tension and centrifugal forces (second term) and the twist-flow coupling (third term). The wave energy equation \eqref{integrated_energy} was derived by \cite{Taroyan2021}, and is valid both in the linear and non-linear regimes.

\section{Numerical analysis of twist evolution}\label{Analysis}
In this work, we investigate the spatio-temporal evolution of small-amplitude twists along a stratified magnetic thread in the presence of plasma inflow. The twists are driven at the footpoints by random photospheric motion. We carry out the investigation as a driven problem to mimic the random photospheric twisting motions, whilst the inflow aims to simulate plasma accumulation along the thread.
The inflows considered in this paper are strictly sub-Alfv\'enic with much lower flow
speeds than the Alfv\'en speed. \cite{Sumner2020} investigated setups that yield super-Alfv\'enic inflows along a uniform thread. These setups lead to the formation of amplifying twists with steep gradients in the neighbourhood of the critical points, where the flow and Alfv\'en speeds become equal to one another. 

We previously examined an initial value problem for a uniform thread with footpoints rooted in the photosphere. A vanishing boundary condition on the azimuthal velocity was imposed to ensure no influx
of azimuthal kinetic energy. A boundary condition on the magnetic field perturbation follows from the momentum equation. For flow speeds much lower than the Alfv\'en speed, this second condition expresses the continuity of the azimuthal magnetic field. 

The driven boundary conditions we utilise here are similar, but they now require the incorporating of an additional random driver:
\begin{equation}\label{eq_driven_boundaries}
\begin{split}
& v_\theta = f^\pm, \\
& \frac{\partial b_\theta}{\partial s} -
\frac{b_\theta}{B_s} \frac{\partial B_s}{\partial s} =
\frac{\mu_0 \rho}{B_s} \frac{\partial f^\pm}{\partial t},
\end{split}
\end{equation}
where $f^\pm$ represents the random velocity driver at the two footpoints. This is described below in more detail. The first condition in \eqref{eq_driven_boundaries} is known to be essential, whereas the second condition is known as a natural boundary condition. The velocity driver mimics random photospheric motions with an azimuthal component. These motions result in twists propagating along the magnetic field. The photospheric boundaries
are located at $s=\pm L/2$.
The boundary condition on the magnetic field perturbation follows from the momentum equation \eqref{Eq_Motion}. However, the second term on the left-hand side of equation \eqref{Eq_Motion} can be ignored and can be simplified to the above condition for small Alfv\'enic Mach numbers to reduce computation time. We applied this simpler condition because the Alfv\'enic Mach numbers used in the setups are generally small. Tests produced negligible differences in outcomes. Alternatively, one may simply require that the $b_\theta$ variable be continuous across the photospheric boundary so that $\displaystyle{\frac{\partial b_\theta}{\partial s}=0}$. The resulting outcomes are again  similar and no significant differences are produced.

\begin{figure}
\centering
\resizebox{\hsize}{!}{\includegraphics{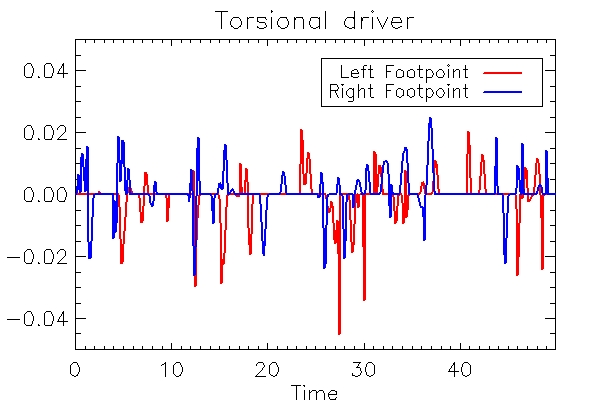}}
\caption{Photospheric motions are modelled as small random buffeting of the footpoints throughout the duration of the simulation. The amplitude of these perturbations is always smaller than $0.05$. In total, 50 pulses are injected at each footpoint, with a uniformly random distribution of durations between $0.05t$ and $1.05t.$}
\label{fig:random_driver}
\end{figure}

Our analysis utilises the Advanced Maths and Statistics Library for IDL, solving the linear partial differential equations using the method of lines approach (IMSL\_PDE\_MOL). We use a spatial grid of 500 points, and a continuous time variable with outputs produced every $0.1$ time steps, where time is normalised with respect to the Alfv\'en travel time, $t_A$ (see section 2.1). Higher spatial resolutions were prohibited because they lead to excessive computation times, although testing showed negligible differences with a 1000-point grid.

The magnetic threads used in this work experience random footpoint motions throughout the simulations. The same driven boundary conditions are used for each setup, and are as shown in Figure \ref{fig:random_driver}. The pulses are injected randomly throughout the duration of the simulation, with amplitudes in arbitrary units of less than $0.05$ and duration between $0.05t$ and $1.05$. These pulses are implemented according to equation \eqref{eq_driven_boundaries}. A number of drivers were generated and tested, with the driver presented here being a representative of the average case.

We present results for three different cases, which represent threads reaching different heights in the atmosphere. Rather than aiming to describe a specific observed phenomenon, such as by prescribing specific contrasts in plasma density and magnetic field strength to describe a solar prominence or chromospheric arch, we instead study the parameter regimes appropriate to a variety of conditions.

\subsection{Background conditions}\label{Sec_BC}
The evolution of the thread is investigated for a duration of 50 normalised time units. The contrast between the initial and final plasma densities at the thread centre depends on the value of the time scale $t_0$, which also determines the inflow speed. In what follows, we have taken a value of $t_0=1$ corresponding to a contrast of around 50 between the final and initial densities at the centre. 

The model described in Section \ref{Model} contains three main length scales: $\sigma_1$, $\sigma_2$, and $\sigma_3$. The parameter $\sigma_1$ determines the initial density contrast, whilst $\sigma_3$ prescribes the background magnetic field gradient in \eqref{Eq_Magnetic}. The parameter $\sigma_2$ is used to control the width of the central accumulation of plasma. All background quantities are normalised with respect to their initial values at the footpoints, meaning that the outcomes in different cases have similar photospheric values and can be compared. 

We note that all three cases presented below have the same thread length of $1$ in normalised dimensionless units. However, differences due to the the scale lengths, $\sigma_1$ and $\sigma_3$,  will result in different thread lengths in dimensional units. 
\subsubsection{Short thread}
\begin{figure}
\centering
\resizebox{\hsize}{!}{\includegraphics{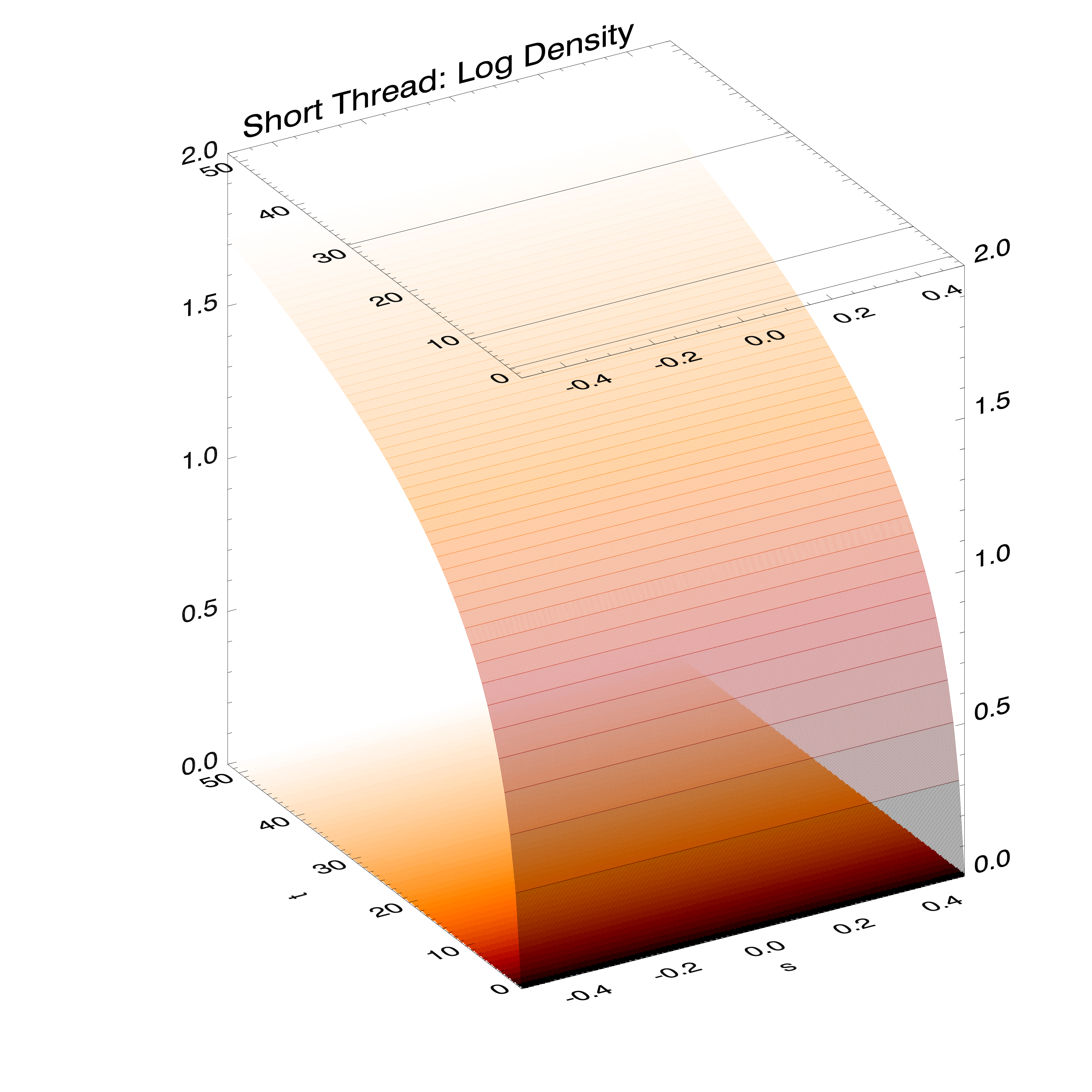}}
\resizebox{\hsize}{!}{\includegraphics{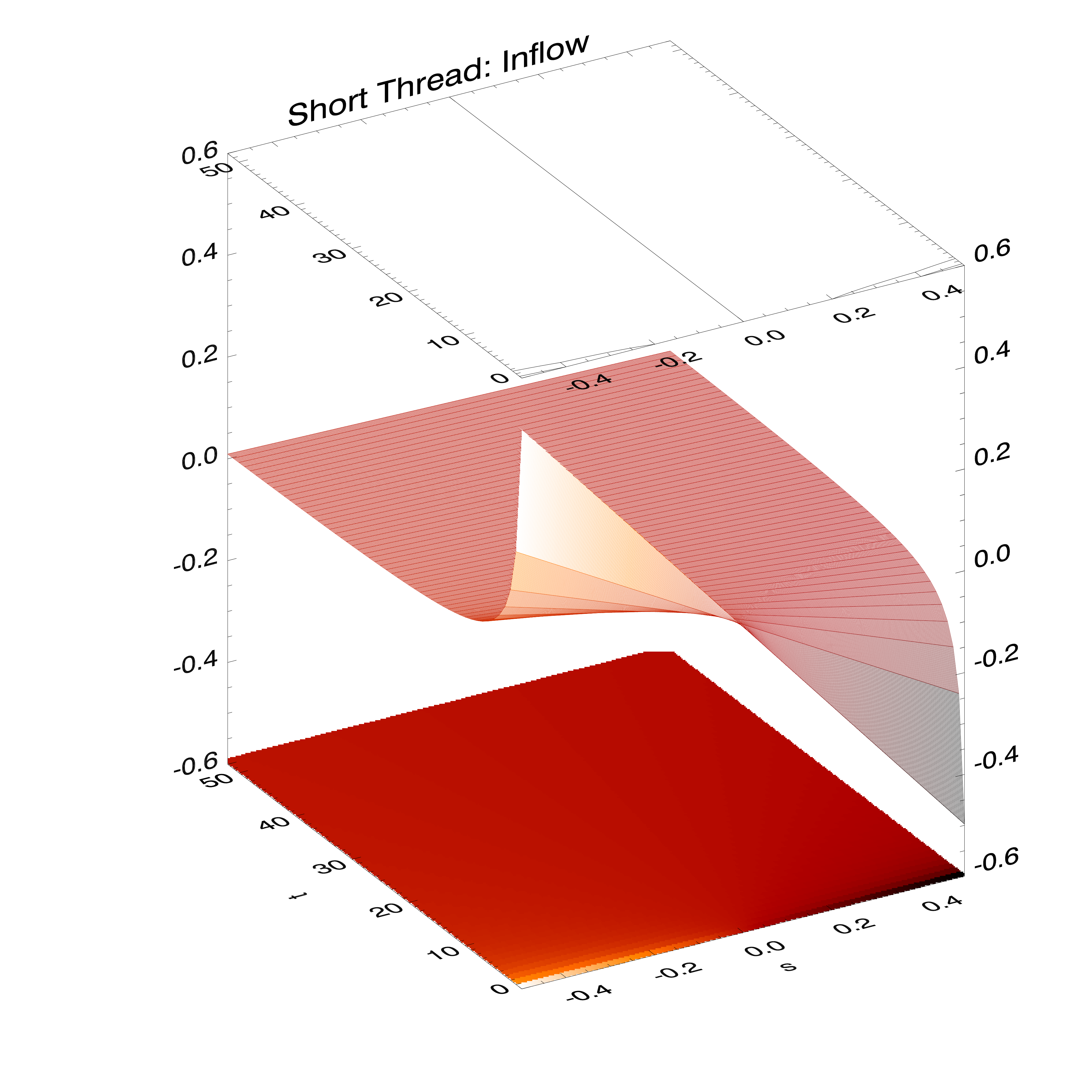}}
\caption{Background conditions for the short thread setup. 
Each plot combines an image (bottom), a surface plot (middle), and a contour plot (top) of the corresponding quantity.
\textbf{Top:} Logarithmic density profile along the simulated thread over time. 
\textbf{Bottom:} Associated inflow profile.}
\label{fig:Low_Density_background}
\end{figure}
Using the introduced length scales, $\sigma_i$, here we present three representative case studies. The first case describes a short thread, where very large $\sigma_i$ values are used. This case also serves to validate the new model by simulating nearly uniform conditions present in \cite{Sumner2020}. This short-thread case describes nearly uniform conditions low in the solar atmosphere. The values of the length scales, $\sigma_i$, used to produce this setup and the corresponding initial contrasts between the thread footpoint or base (denoted by index $b$) and the centre (denoted by index $c$) are as follows:
\begin{align*} 
\sigma_1 = 10.00 &\longrightarrow \frac{\rho_b}{\rho_c} = \frac{100}{99}, \\ 
\sigma_2 = 10.00 &\longrightarrow \text{ nearly uniform density increase}, \\ 
\sigma_3 = 10.00 &\longrightarrow \frac{B_b}{B_c} = \frac{100}{99}. 
\end{align*}
The logarithmic density plot and the associated inflow profile for this setup are shown in Figure \ref{fig:Low_Density_background}. The process of plasma filling occurs uniformly rather than through accumulation at the thread centre. The inflow speed has a negative gradient throughout the thread and reaches its maximum values at the footpoints. 
\subsubsection{Long thread}
\begin{figure}
\centering
\resizebox{\hsize}{!}{\includegraphics{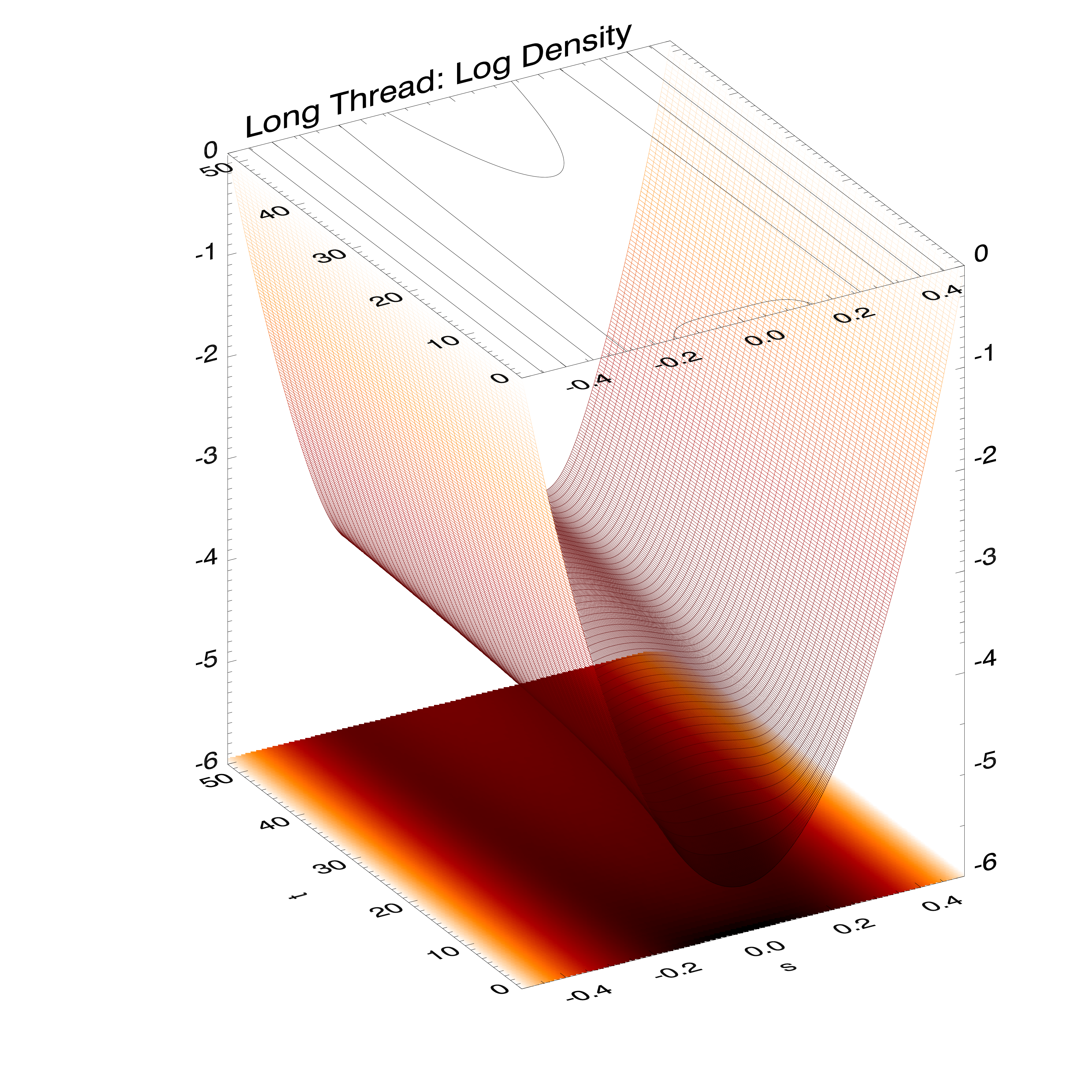}}
\resizebox{\hsize}{!}{\includegraphics{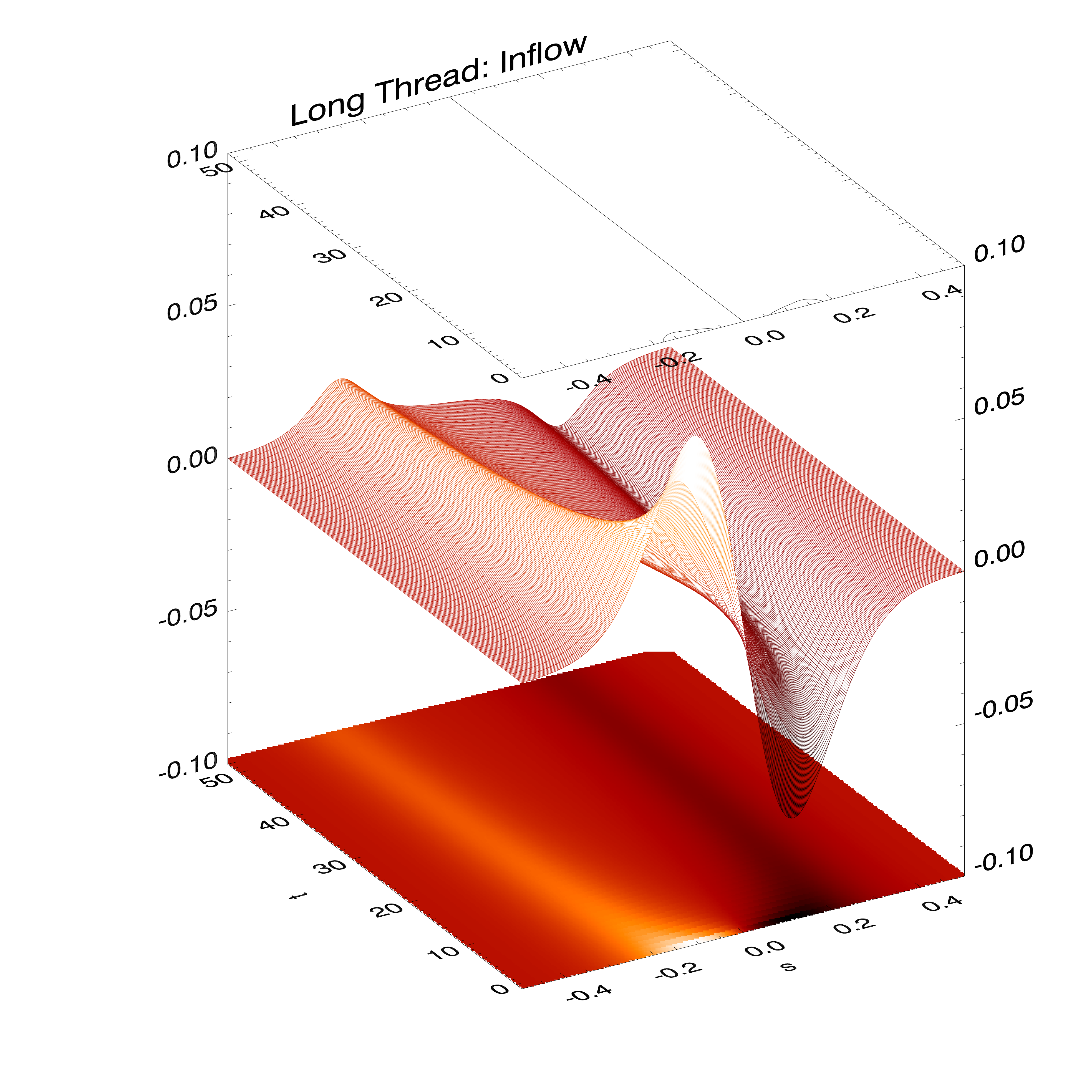}}
\caption{ Background conditions for the long-thread setup. Each plot combines an image (bottom), a surface plot (middle),
and a contour plot (top) of the corresponding quantity. Top: Logarithmic density profile along the simulated thread over time.
Bottom: Associated inflow profile. }
\label{fig:High_Density_background}
\end{figure}
We secondly make use of a long-thread setup, which is intended to represent the kind of gradient that may be present along a thread extending from the photosphere into the corona. This setup uses the following values for the scale lengths, $\sigma_i$, and represents the following contrast between the footpoints and the centre of the thread:
\begin{align*} 
\sigma_1 = 0.14 &\longrightarrow \frac{\rho_b}{\rho_c} = 3.46\times 10^5, \\ 
\sigma_2 = 0.20 & \longrightarrow \text{condensation length scale of $0.2 \times L$,} \\ 
\sigma_3 = 0.20 &\longrightarrow \frac{B_b}{B_c} =518.
\end{align*}
The logarithmic density plot and the associated inflow profile for this setup are shown in Figure \ref{fig:High_Density_background}. The density and the magnetic field contrasts are the highest for the long thread case. 
Unlike the previous case of a short uniform thread, the inflow is no longer highest at the footpoints. Similarly, the gradient of the inflow velocity is no longer negative throughout the thread. 
\subsubsection{Medium thread}

\begin{figure}
\centering
\resizebox{\hsize}{!}{\includegraphics{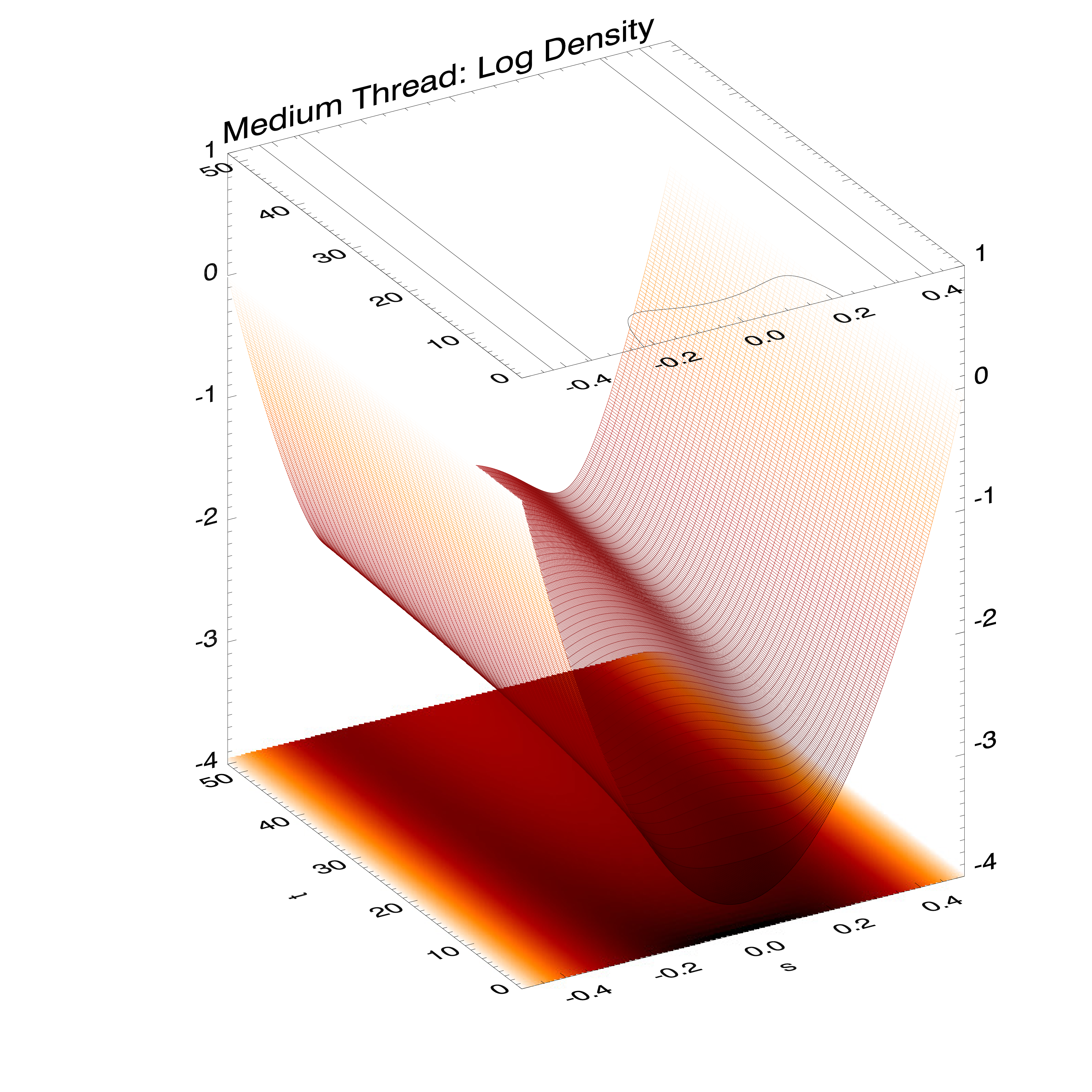}}
\resizebox{\hsize}{!}{\includegraphics{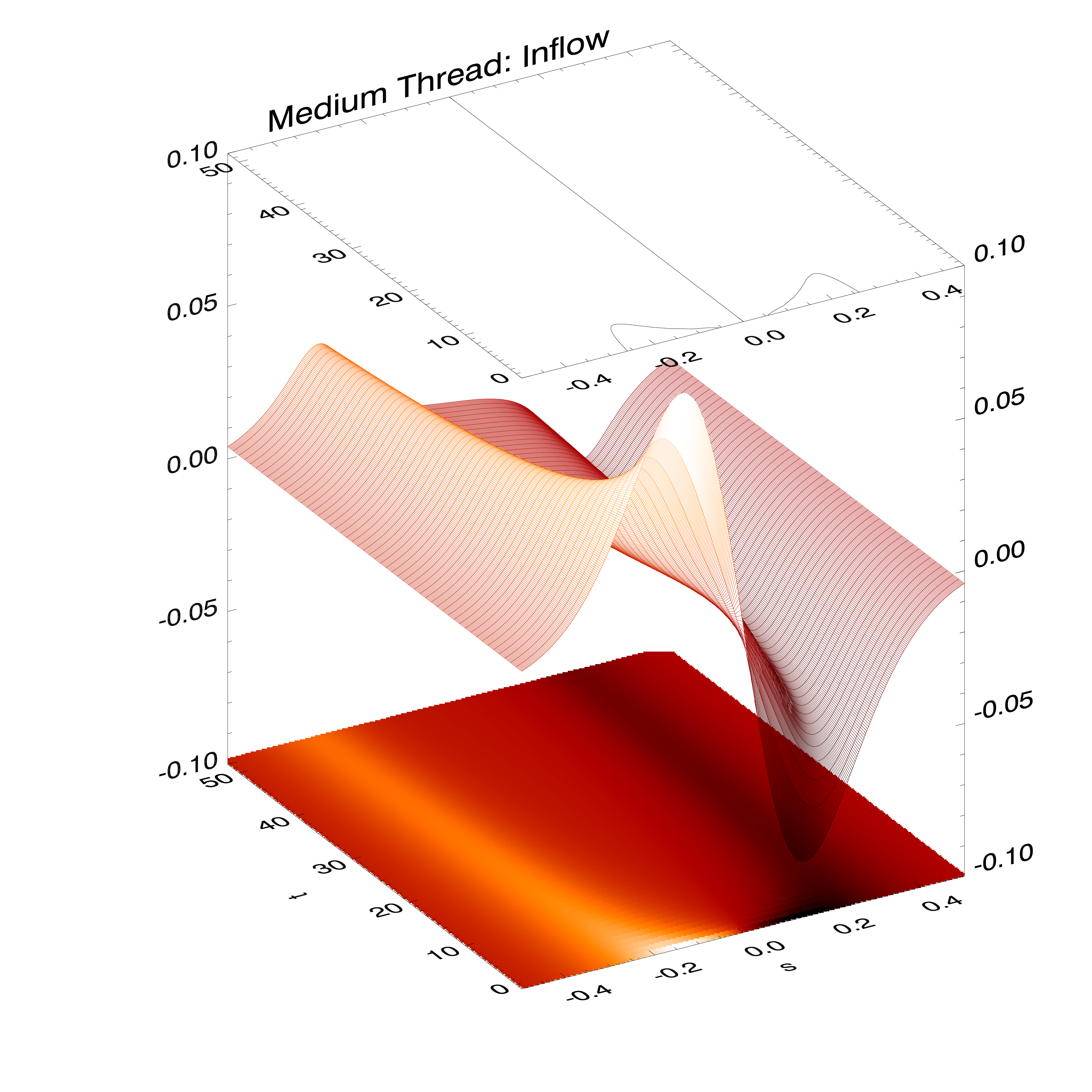}}
\caption{Background conditions for the medium-thread setup. Each plot combines an image (bottom), a surface plot (middle), and a contour plot (top) of the corresponding quantity. Top: Logarithmic density profile along the simulated thread over time. Bottom: Associated inflow profile.}
\label{fig:Med_Density_background}
\end{figure}
Finally, we use a mid-range density contrast, describing a thread extending from the photosphere to the chromosphere. This medium-length thread uses the following $\sigma$ values to represent the following value contrast between the footpoints and the centre of the thread:
\begin{align*} 
\sigma_1 = 0.17 &\longrightarrow \frac{\rho_b}{\rho_c} = 5.7\times 10^3, \\ 
\sigma_2 = 0.20 &\longrightarrow \text{ condensation length scale of $0.2 \times L$}, \\ 
\sigma_3 = 0.22 &\longrightarrow \frac{B_b}{B_c} = 175.
\end{align*}
The logarithmic density plot and the associated inflow profile for this setup are shown in Figure \ref{fig:Med_Density_background}. Compared with the previous case of a long thread, the density and the magnetic field contrasts between the photosphere and chromosphere are smaller. The plasma density at the centre of the thread increases by a factor of 50. For simplicity, the scale length of the plasma accumulation is the same as in the previous case. A comparison between Figures \ref{fig:High_Density_background} and \ref{fig:Med_Density_background} shows that the difference between the inflow profiles for the two cases is less pronounced.

\begin{figure}
\centering
\resizebox{\hsize}{!}{\includegraphics{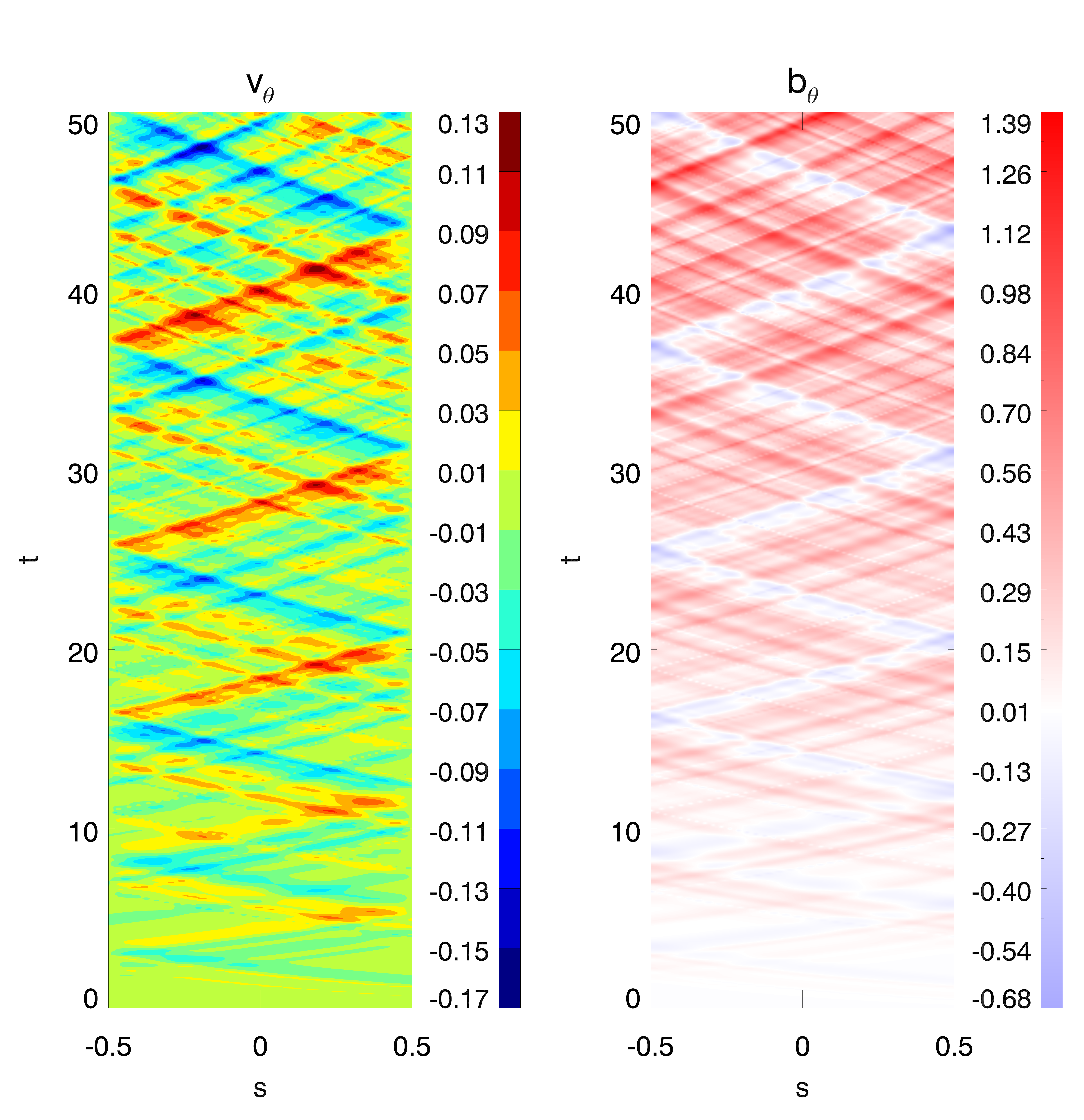}}
\resizebox{\hsize}{!}{\includegraphics{100_100_100_energy-eps-converted-to.pdf}}
\caption{Simulation results for the short-thread setup.  \textbf{Top:} Spatial and temporal evolution of the azimuthal velocity, $v_\theta$, and the magnetic field, $b_\theta$, along the short thread. The perturbations are randomly driven at the footpoints corresponding to $s=\pm\frac{1}{2}$. The corresponding values are indicated with colour bars on the right.  
\textbf{Bottom:} Associated total azimuthal energy, the net energy influx through the boundaries, and magnetic and kinetic energy densities.}
\label{fig:Low_Density_Var_random}
\end{figure}
\begin{figure}
\centering
\resizebox{\hsize}{!}{\includegraphics{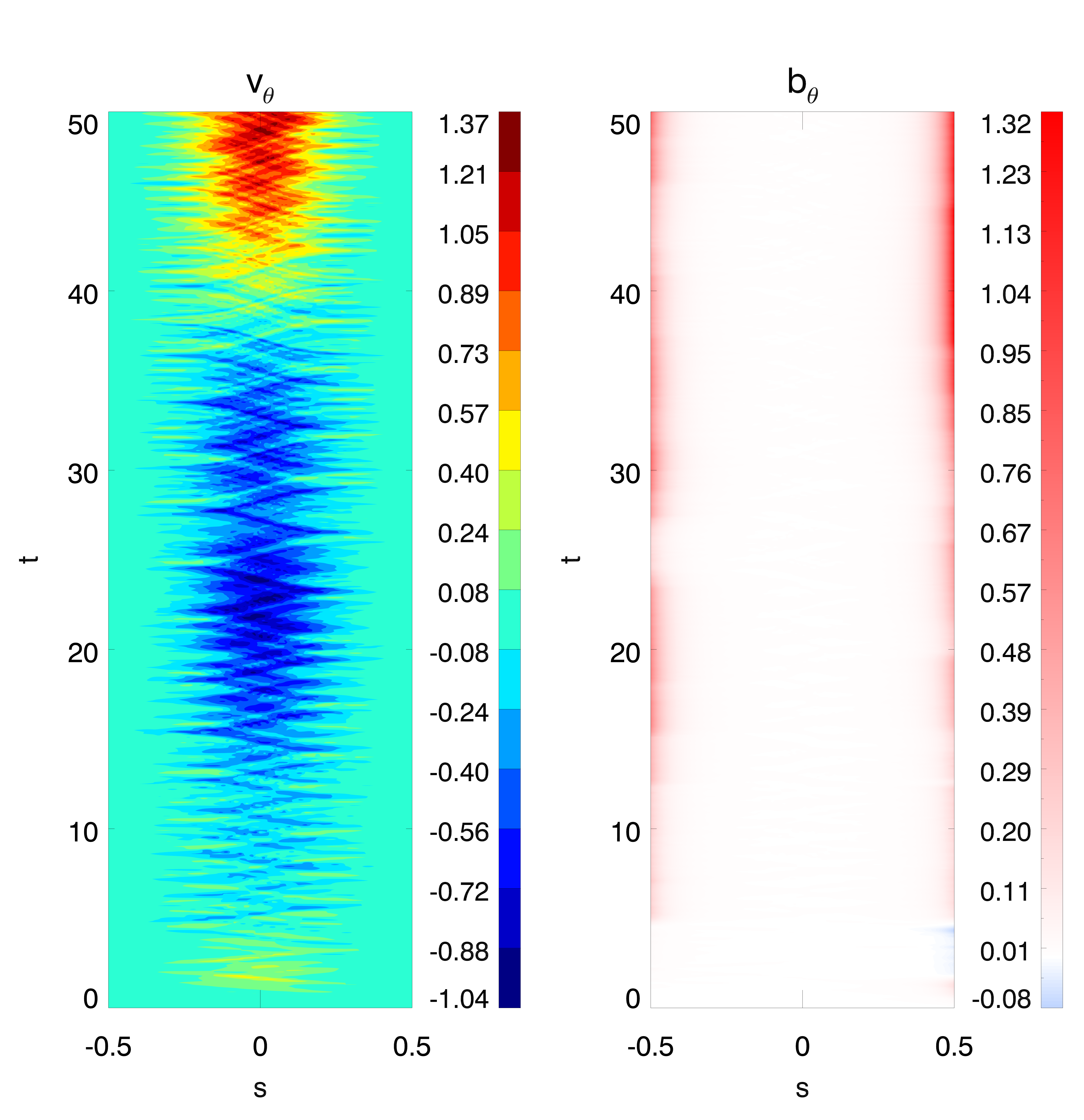}}
\resizebox{\hsize}{!}{\includegraphics{14_20_20_energy-eps-converted-to.pdf}}
\caption{Same as in Figure \ref{fig:Low_Density_Var_random} but for the long-thread setup. }
\label{fig:High_Density_Var_random}
\end{figure}
\begin{figure}
\centering
\resizebox{\hsize}{!}{\includegraphics{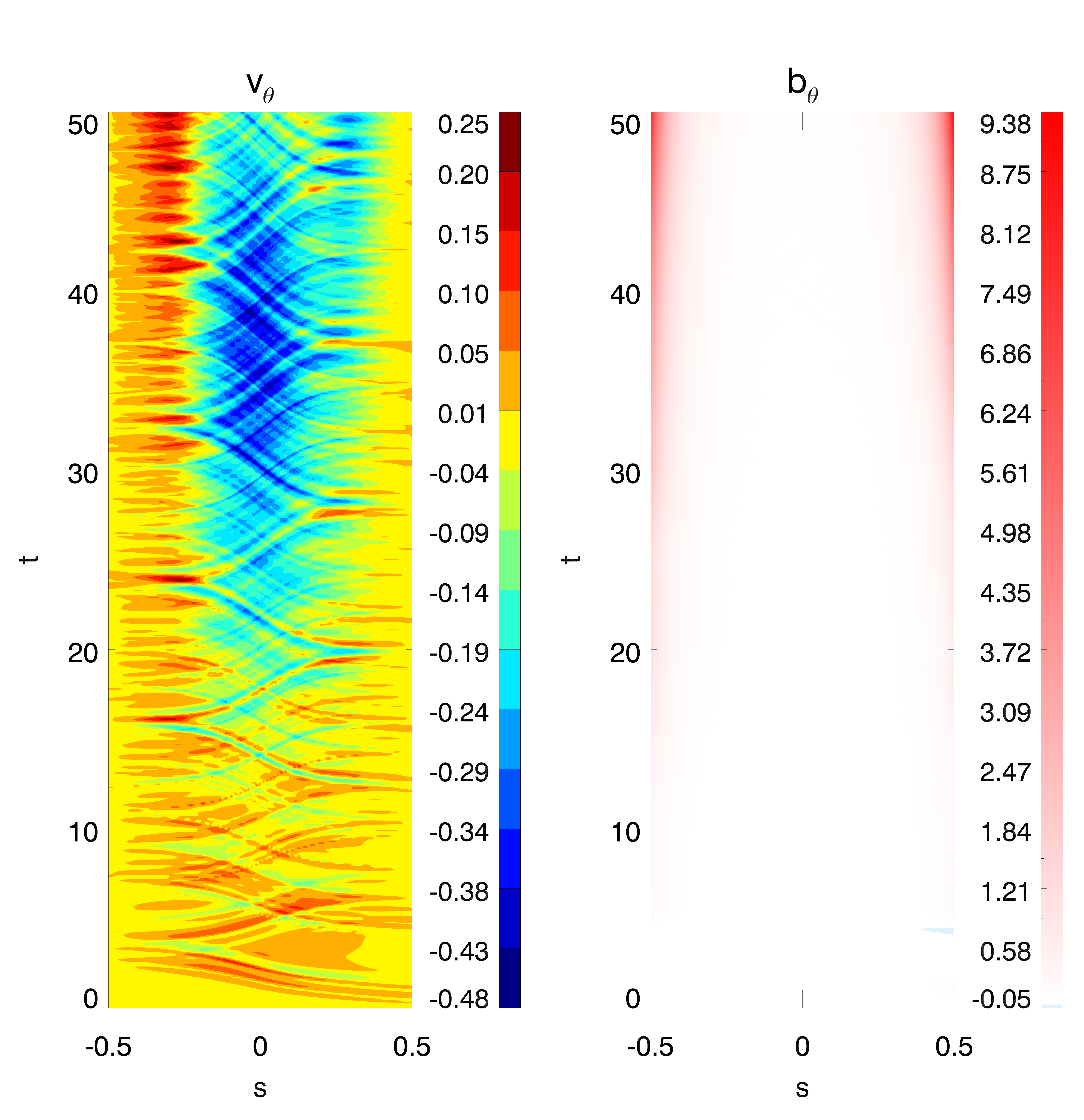}}
\resizebox{\hsize}{!}{\includegraphics{17_20_22_energy-eps-converted-to.pdf}}
\caption{Same as in Figure \ref{fig:Low_Density_Var_random} but for the medium-thread setup. }
\label{fig:Med_Density_Var_random}
\end{figure}

\subsection{Twist evolution}
The next step in our analysis is investigating the evolution of the twists randomly driven at the photospheric footpoints. The analysis is carried out for the three different thread setups   presented above. 

\subsubsection{Short thread}

By the use of sufficiently large length scales, $\sigma_i$, we are able to produce an atmosphere that is mostly flat in density contrast and magnetic field strength (Figure \ref{fig:Low_Density_background}). Such properties would be characteristic of a thread very low in the solar atmosphere. The model is additionally very similar to the uniform background conditions used in our previous study \citep{Sumner2020}. This similarity allows us to validate the new model results and confirm that they are consistent with the previous work. We note that in order to achieve an analytically solvable inflow setup with fixed critical points, a different temporal dependence was selected by \cite{Sumner2020}. 

Figure \ref{fig:Low_Density_Var_random} shows that the results for the short thread are consistent with those in \cite{Sumner2020}. The torsional perturbations are amplified as they travel along the thread, bouncing back and forth at the footpoints. There is continuous amplification of azimuthal magnetic and kinetic energies within the thread throughout the run-time. There is distinct intra-thread amplification of torsional motions, as the total azimuthal energy is amplified, consistently remaining higher than the net azimuthal energy influx (first term on the right-hand side of equation \eqref{integrated_energy}). 

\subsubsection{Long thread}

These conditions yield boundary to apex magnetic field contrast of around $581$. The corresponding density contrast for this setup is $3.46 \times 10^{5}$ between the footpoints and the centre of the thread. The long-thread setup results have three key features. 

The first defining feature is the dominant role of the net azimuthal influx compared with the other two other setups. Figure \ref{fig:High_Density_Var_random} shows very little difference between the curves representing total energy and the net influx. The source terms on the right-hand side of equation \eqref{integrated_energy} have a very minor role. 
Therefore, the present setup is most dependent upon the random driver, and therefore upon the local solar conditions. In some cases, the convergence of azimuthal flux through the driven footpoints leads to moderate amplification similar to that shown in Figure \ref{fig:High_Density_Var_random}. There are cases with a more pronounced amplification due to enhanced net influx. In Figure \ref{fig:High_Density_Var_random} we see the thread initially acting as a sink up until around $t=16$, at which point we see a gradually increasing positive contribution from the source terms to the process of amplification. 

The second important feature we see in all cases with the long-thread setup is an amplifying standing wave. The standing wave is seen both in the velocity with a pronounced antinode at the centre of the thread and in the magnetic field that has antinodes at the footpoints. These motions are seen to amplify during the course of the simulation. The standing wave can be seen in the top left panel of Figure \ref{fig:High_Density_Var_random}. The standing wave feature is also present in the absence of an inflow and the oscillation period remains constant. The period increases when the perturbations amplify due to the introduction of an inflow. The period increase can be seen in Figure \ref{fig:High_Density_Var_random} where the velocity changes sign at $t = 5$, and then at $t = 40$. 

Finally, linked to the second feature is the confinement of the azimuthal magnetic energy to the footpoints of the thread. The azimuthal magnetic field is significantly weaker away from the footpoints. 

The behaviour in the randomly driven case presented here is representative of tests using other random drivers where all three features consistently appear. Further testing with non-driven footpoints and an initial torsional pulse symmetrically injected at the centre of the thread shows that the total azimuthal energy decays over time. This confirms that the growth seen for the long thread setup is mainly caused by the asymmetric injection of pulses and the corresponding net azimuthal influx through the footpoints. This type of asymmetry would be difficult to avoid considering the random nature of the photospheric motions.

\subsubsection{Medium thread}

In order to mimic a thread extending from the photosphere to the chromosphere, we choose the following parameter values for the background: a boundary-to-apex magnetic field contrast of approximately 175, and an initial density contrast of around $5.7 \times 10^3$ between the footpoints and the centre of the thread.

Figure \ref{fig:Med_Density_Var_random} shows the evolution of the azimuthal velocity, magnetic field, and the corresponding energies when the footpoints are driven randomly. The top panels display a quasi-standing wave feature that we have already seen for the long-thread setup. The period increase associated with the amplitude increase is more pronounced due to more rapid growth. 

The increasing amplitude is clearly seen in $b_\theta$ along the boundaries of the thread, and in $v_\theta$ around the centre of the thread. 
The total azimuthal energy in the system always exceeds the net energy influx and therefore an additional source of energy is present. This additional energy is caused by the positive balance of the source terms on the right hand side of the energy equation \ref{integrated_energy}. In testing other driver setups, this result is always seen, with significantly higher amplitudes in some cases, whilst the results presented here are more typical of the average. 

The bottom panel of Figure \ref{fig:Med_Density_Var_random} shows that the growth in the medium-thread case is primarily due to the increase in magnetic energy.
A comparison of Figure \ref{fig:Med_Density_Var_random} with Figures \ref{fig:Low_Density_Var_random} and \ref{fig:High_Density_Var_random} reveals that the amplification of the torsional perturbations is most efficient in the medium thread. We discuss the reason for this rapid growth in the following section. 

Our choice of $\sigma_3=0.22$ here is intended as a lower magnetic field strength contrast than for long threads, but is also chosen to restrain amplification. Strong magnetic field contrasts lead to much greater amplification. For example, choosing $\sigma_3=0.20$ resulted in such rapid growth that the gradients exceeded the error limits on the simulation in the linear regime.

\section{Discussion}\label{Discussion}
We have mainly considered the dependence of the twist amplification on the length of the thread along which twists propagate. Length is normalised with respect to the thread length, and therefore it is implicitly considered in the density and magnetic field length scales, $\sigma_1$ and $\sigma_3$. Long threads are represented by small length scales, whereas short threads are represented by large length scales. 

The length scale, $\sigma_2$, and the timescale, $t_0$, of the plasma accumulation have similar effects on the evolution of the perturbations. Increasing the length scale or reducing the timescale of the plasma accumulation leads to more rapid growth and vice versa. Expression \eqref{Eq_Inflow} shows that these changes will lead to higher inflow velocities with stronger gradients which are the key ingredients in the amplification process.

Our results show growth of torsional perturbations regardless of thread length. In what follows we demonstrate that these perturbations are amplified in different ways depending on the thread length. The mechanisms are explained and discussed separately for the three setups. 

In order to understand the details of twist evolution, we consider the energy equation \eqref{InflowBCEnergy}. For a short thread, there is no variation in the background magnetic field, and therefore the first source term on the right-hand side will vanish. The amplification is therefore caused by the combined influence of a net influx of energy through the footpoints and by the wave-flow coupling along the thread. 

The latter effect is represented by the second source term on the right-hand side of equation \eqref{InflowBCEnergy}. It remains positive throughout the thread because of the negative flow gradient and a constant magnetic field, and therefore provides an additional source of energy at all points within the thread. Figure \ref{fig:Low_Density_Var_random} shows that total azimuthal energy always remains higher than the net energy influx because of the presence of this additional energy source.

Energy is extracted from the inflow and converted into torsional motions. The back-reaction of the amplified torsional perturbations on the inflow will only emerge in the non-linear stage of evolution. The results for the short thread are analogous to those presented by \cite{Sumner2020}: the random pulses amplify as they travel along the thread. 

The situation is different for medium and long threads where variations in the background density and magnetic field introduce additional effects. Firstly, as indicated by Figures \ref{fig:Med_Density_background} and \ref{fig:High_Density_background}, the flow gradient changes sign, and therefore the source term discussed above may no longer purely represent a source of energy. Instead it acts as a sink in regions of positive flow gradient and a source of energy in regions of negative flow gradient. Secondly, the remaining source terms in the energy equation are no longer zero because of variations in the background magnetic field and their role needs to be clarified.

We note that for medium and long threads, the magnetic field perturbation is highest near the footpoints. We also note that the chosen length scale of plasma accumulation, $\sigma_2$, is small, and so the density near the footpoints remains almost unchanged: $\frac{\partial \rho}{\partial t}\approx 0$. Therefore, from the continuity equation \eqref{Eq_Continuity} we have:
\begin{equation}\label{continuity_mod}
\frac{\partial \ln{\rho}}{\partial s} + \frac{\partial}{\partial s} \ln\left|\frac{U}{B_s} \right| \approx 0
\end{equation}
near the thread footpoints. Combining the above relation \eqref{continuity_mod} and the solenoidal condition condition \eqref{solenoidal}, we can rewrite the energy equation \eqref{InflowBCEnergy} in the following form:
\begin{equation}\label{Energy_mod} 
\frac{\partial}{\partial t}\left(\frac{W_T}{B_s}\right) + \frac{\partial }{\partial s} \left( \frac{F_W}{B_s} \right) \approx \frac{U}{B_s}\left( \frac{\partial \ln \rho}{\partial s} W_m - \frac{\partial \ln B_s}{\partial s}\left[\frac{b_\theta ^2}{\mu_0} - \rho v_\theta ^2 \right]\right).
\end{equation}
The source terms on the right-hand side represent the wave-flow coupling, and the difference between the tension and the centrifugal forces. Away from the footpoints, the magnetic energy, $W_m$, is small, and therefore the source terms on the right-had side of equation \eqref{Energy_mod} are dominated by the centrifugal force.

Equation \eqref{Energy_mod} demonstrates that the first term is negative near the footpoints of medium and long threads, and therefore its main role is to extract energy from the torsional perturbations. This effect is most pronounced in the long thread with strong density gradients near the footpoints. The medium thread experiences a similar effect, but it is less significant because the density gradient is less steep. Therefore, the results of our investigation show that the wave-flow coupling represented by the first term on the right-hand side of equation \eqref{Energy_mod} is capable of acting as both a sink and a source for torsional energy within a thread, depending on the thread length and the corresponding density profile.

In contrast to the wave-flow coupling term, the centrifugal and tension forces only appear when there is a gradient in the background magnetic field. The tension force acts as a source of energy and the centrifugal force represents a sink for the torsional perturbations. The overall effect is determined by the difference between the two forces along the thread. The tension and centrifugal forces are essentially determined by the magnetic and kinetic energies of the perturbations, correspondingly. 

The total magnetic and kinetic energies, $W_m$ and $W_k$, along the threads are plotted in the bottom panels of Figures \ref{fig:Low_Density_Var_random}, \ref{fig:High_Density_Var_random}, and \ref{fig:Med_Density_Var_random}. The difference between the two forces plays no role in a short thread where the amplification mechanism is different. However, in a medium thread there is a growing difference between the tension and the centrifugal forces as the perturbations amplify. This leads to very rapid growth in medium threads where the magnetic energy is higher than the net influx and almost overlaps with the total energy. Equation \eqref{Energy_mod} shows that a strong gradient in the background magnetic field facilitates the process of amplification.

The source terms balance each other in long threads, and the main source of amplification is the net influx of azimuthal energy through the footpoints. The influx is a natural consequence of the asymmetry between the drivers at the two footpoints and it can only be avoided when the two drivers are perfectly symmetric throughout the simulations.   

The bottom panel of Figure \ref{fig:High_Density_Var_random} shows that any temporary increase in the centrifugal force is associated with damping or reduced amplification. Equation \eqref{Energy_mod} shows that a strong density gradient near the footpoints of long loops will inhibit the process of amplification.

In our previous paper, setups were tested within which regions of super-Alfv\'enic flows led to the confinement of the perturbations at critical points, causing localised sharp growths in torsional energy. This effect does not occur in the present work, which has stratified background quantities. The Alfv\'en speed is always higher than the inflow speed and therefore no regions of super-Alfv\'enic flows are present.

Our results indicate that threads extending to chromospheric heights are most susceptible to twisting through the presented mechanism. However, we have not yet addressed their possible role in prominence formation. A well-known scenario suggests that localised heating above the footpoints leads to the evaporation of chromospheric material, which is carried along field lines. Accumulation of evaporated material causes a rise in density, enhanced radiation, and cooling. This method of mass accumulation and condensation has been modelled in the literature \citep{Gibson2018}. 

A well-known model that could support the dense prominence plasma was suggested by \cite{Kippenhahn1957}: the magnetic tension force due to bent field lines provides support against the force of gravity. On the other hand, an increasing magnetic pressure from the centre of the prominence compresses the plasma to oppose the outward gas pressure gradient.

Our analysis is carried out in the linear regime and shows the formation of a strongly twisted magnetic field near the footpoints. In the non-linear stage of evolution, the plasma will experience a ponderomotive force that is expressed through the gradient of $b_\theta ^2$. The ponderomotive force will be acting from the footpoints towards the centre and could provide support against the gravitational force and the gas pressure gradient. However, the likelihood of this mechanism leading to prominence formation, similar to the \cite{Kippenhahn1957} model, must be verified through a detailed analysis in the non-linear regime. Another interesting consequence could be the role of the twist amplification mechanism in the heating and cooling cycle within coronal loops that are associated with inflows and outflows.

\section{Conclusions}\label{Conclusions}
We investigated the temporal and spatial evolution of linear torsional Alfvénic perturbations along a simulated magnetic thread, in conditions representative of a time-dependent stratified solar atmosphere. The perturbations are randomly driven at the photospheric footpoints. We demonstrate that magnetic threads may become twisted by three distinct mechanisms across different scales. We conclude that a negative gradient between the lower and upper parts of the atmosphere inhibits the process of amplification. In contrast, a similar gradient in the background magnetic field acts to facilitate the process of amplification.

Short threads with uniform conditions support torsional waves that become amplified due to azimuthal energy influx through the footpoints and through coupling with the decelerating inflow of plasma along the thread. The results produced by \cite{Sumner2020} are replicated as a control case in this work, in Figure \ref{fig:Low_Density_Var_random}, and are extended in light of the new results.

The incorporation of a variable magnetic field along the thread introduces centrifugal and magnetic tension forces which were previously absent. The tension force has a positive effect whereas the centrifugal force and the wave-flow coupling have a negative impact on the process of twist amplification.

The process of amplification is most efficient in threads of medium length that extend to chromospheric heights. This is due to the combined action of the tension force and the energy influx. Long threads that reach coronal heights are amplified less efficiently because of steeper density gradients. 

Another key finding of this work is that regardless of the parameter values, a standing wave is set up along the thread in cases with a stratified atmosphere. This is most prominent in long threads that reach coronal heights and occurs regardless of the form of the random buffeting motions at the footpoints. Contour plots show formation of velocity, $v_\theta$, nodes at the footpoints and a magnetic field, $b_\theta$, node at the thread centre. The antinodes in $b_\theta$ are associated with the strong twisting of the field lines near the footpoints.

Non-linear modelling should be carried out to further describe the evolution of the amplified twists and their possible consequences. These include the formation of a twisted prominence topology on a global scale, tornado like motions, and prominence eruptions.
\begin{acknowledgements}
Chloe Sumner would like to thank the STFC for their financial support (ST/S505225/1).
\end{acknowledgements}
\bibliographystyle{aa}
\bibliography{Bibliography}

\begin{thebibliography}{26}
\expandafter\ifx\csname natexlab\endcsname\relax\def\natexlab#1{#1}\fi

\bibitem[{Arregui {et~al.}(2018)Arregui, Oliver, \& Ballester}]{Arregui2018}
Arregui, I., Oliver, R., \& Ballester, J. 2018, Living Reviews in Solar
  Physics, 15

\bibitem[{{Aschwanden} \& {Schrijver}(2002)}]{Aschwanden2002}
{Aschwanden}, M.~J. \& {Schrijver}, C.~J. 2002, \apjs, 142, 269

\bibitem[{{Gibson}(2018)}]{Gibson2018}
{Gibson}, S.~E. 2018, Living Reviews in Solar Physics, 15, 7

\bibitem[{{Hillier} {et~al.}(2013){Hillier}, {Morton}, \&
  {Erd{\'e}lyi}}]{Hillier2013}
{Hillier}, A., {Morton}, R.~J., \& {Erd{\'e}lyi}, R. 2013, \apjl, 779, L16

\bibitem[{{Hollweg}(1992)}]{Hollweg1992}
{Hollweg}, J.~V. 1992, \apj, 389, 731

\bibitem[{{Hollweg} {et~al.}(1982){Hollweg}, {Jackson}, \&
  {Galloway}}]{Hollweg1982}
{Hollweg}, J.~V., {Jackson}, S., \& {Galloway}, D. 1982, \solphys, 75, 35

\bibitem[{Jess {et~al.}(2009)Jess, Mathioudakis, Erdélyi, Crockett, P~Keenan,
  \& J~Christian}]{Jess2009}
Jess, D., Mathioudakis, M., Erdélyi, R., {et~al.} 2009, Science (New York,
  N.Y.), 323, 1582

\bibitem[{{Kippenhahn} \& {Schl{\"u}ter}(1957)}]{Kippenhahn1957}
{Kippenhahn}, R. \& {Schl{\"u}ter}, A. 1957, \zap, 43, 36

\bibitem[{{Li} {et~al.}(2012){Li}, {Morgan}, {Leonard}, \&
  {Jeska}}]{Morgan2012}
{Li}, X., {Morgan}, H., {Leonard}, D., \& {Jeska}, L. 2012, \apjl, 752, L22

\bibitem[{{Liggett} \& {Zirin}(1984)}]{Liggett1984}
{Liggett}, M. \& {Zirin}, H. 1984, \solphys, 91, 259

\bibitem[{{Liu} {et~al.}(2019){Liu}, {Nelson}, {Snow}, {Wang}, \&
  {Erd{\'e}lyi}}]{2019NatCo..10.3504L}
{Liu}, J., {Nelson}, C.~J., {Snow}, B., {Wang}, Y., \& {Erd{\'e}lyi}, R. 2019,
  Nature Communications, 10, 3504

\bibitem[{Luna {et~al.}(2012)Luna, Karpen, \& DeVore}]{Luna_2012}
Luna, M., Karpen, J.~T., \& DeVore, C.~R. 2012, The Astrophysical Journal, 746,
  30

\bibitem[{Luna {et~al.}(2015)Luna, Moreno-Insertis, \& Priest}]{Luna_2015}
Luna, M., Moreno-Insertis, F., \& Priest, E. 2015, The Astrophysical Journal,
  808, L23

\bibitem[{Parenti(2014)}]{Parenti2014}
Parenti, S. 2014, Living Reviews in Solar Physics, 11, 1

\bibitem[{{Price} {et~al.}(2015){Price}, {Taroyan}, {Innes}, \&
  {Bradshaw}}]{Price2015}
{Price}, D.~J., {Taroyan}, Y., {Innes}, D.~E., \& {Bradshaw}, S.~J. 2015,
  \solphys, 290, 1931

\bibitem[{Srivastava {et~al.}(2017)Srivastava, Shetye, Murawski, Doyle,
  Stangalini, Scullion, Ray, Wójcik, \& Dwivedi}]{Srivastava2017}
Srivastava, A.~K., Shetye, J., Murawski, K., {et~al.} 2017, Scientific Reports,
  7

\bibitem[{Su {et~al.}(2012)Su, Wang, Veronig, Temmer, \& Gan}]{Su_2012}
Su, Y., Wang, T., Veronig, A., Temmer, M., \& Gan, W. 2012, The Astrophysical
  Journal, 756, L41

\bibitem[{{Sumner} \& {Taroyan}(2020)}]{Sumner2020}
{Sumner}, C. \& {Taroyan}, Y. 2020, \aap, 642, A181

\bibitem[{{Taroyan} {et~al.}(2006){Taroyan}, {Bradshaw}, \&
  {Doyle}}]{Taroyan2006}
{Taroyan}, Y., {Bradshaw}, S.~J., \& {Doyle}, J.~G. 2006, \aap, 446, 315

\bibitem[{Taroyan {et~al.}(2021)Taroyan, Hovhannisyan, \& Sumner}]{Taroyan2021}
Taroyan, Y., Hovhannisyan, G., \& Sumner, C. 2021, Monthly Notices of the Royal
  Astronomical Society: Letters, 506, L64

\bibitem[{{Taroyan} \& {Soler}(2019)}]{Taroyan_2019}
{Taroyan}, Y. \& {Soler}, R. 2019, A\&A, 631, A144

\bibitem[{{Vernazza} {et~al.}(1981){Vernazza}, {Avrett}, \&
  {Loeser}}]{Vernazza1981}
{Vernazza}, J.~E., {Avrett}, E.~H., \& {Loeser}, R. 1981, \apjs, 45, 635

\bibitem[{Wedemeyer {et~al.}(2013)Wedemeyer, Scullion, van~der Voort, Bosnjak,
  \& Antolin}]{Wedemeyer_2013}
Wedemeyer, S., Scullion, E., van~der Voort, L.~R., Bosnjak, A., \& Antolin, P.
  2013, The Astrophysical Journal, 774, 123

\bibitem[{{Wedemeyer-B{\"o}hm} {et~al.}(2012){Wedemeyer-B{\"o}hm}, {Scullion},
  {Steiner}, {Rouppe van der Voort}, {de La Cruz Rodriguez}, {Fedun}, \&
  {Erd{\'e}lyi}}]{Bohm2012}
{Wedemeyer-B{\"o}hm}, S., {Scullion}, E., {Steiner}, O., {et~al.} 2012, \nat,
  486, 505

\bibitem[{{Williams} \& {Taroyan}(2018)}]{Williams2018}
{Williams}, T. \& {Taroyan}, Y. 2018, \apj, 852, 77

\bibitem[{Xia \& Keppens(2016)}]{Xia2016}
Xia, C. \& Keppens, R. 2016, The Astrophysical Journal, 823, 22

\end{thebibliography}
\begin{appendix}
\section{Derivation of the Induction Equation \eqref{Eq_Induction}}
The azimuthal component of the induction equation for axisymmetric motions in curviliear coordinates reads \citep{Hollweg1982, Taroyan2021}:
\begin{equation}\label{A1}
\frac{\partial}{\partial t} \left( \frac{b_\theta }{ r B_s} \right)+ 
\frac{\partial}{\partial s} \left( \frac{b_\theta}{rB_s} U \right)=
\frac{\partial}{\partial s} \left( \frac{v_\theta }{ r}\right).    
\end{equation}
We note that for field lines close to the symmetry axis the solenoidal condition \eqref{solenoidal} derived by \cite{Taroyan2021} is applicable. It is different from the one proposed by \cite{Hollweg1982} that is applicable to a radially expanding field. We apply condition \eqref{solenoidal} to equation \eqref{A1} to obtain the induction equation \eqref{Eq_Induction}:
\begin{equation}\label{A2}
\frac{\partial b_\theta}{\partial t} + 
\frac{\partial}{\partial s} \left( b_\theta U \right)=
\frac{\partial}{\partial s} \left( B_s v_\theta \right),
\end{equation}
\section{Derivation of the momentum equation \eqref{Eq_Motion}}
The azimuthal component of the momentum equation reads \citep{Hollweg1982, Taroyan2021}:
\begin{equation}\label{Eq_B1}
\frac{\partial}{\partial t} \left( \frac{r \rho v_\theta }{ B_s} \right) + \frac{\partial}{\partial s} \left( \frac{r\rho v_\theta}{B_s} U \right) = \frac{1}{\mu_0}\frac{\partial}{\partial s} \left( r b_\theta \right),    
\end{equation}
We rewrite it in the following form:
\begin{equation} \label{Eq_B2}
\frac{\rho r}{B_s}\frac{\partial v_\theta}{\partial t} + \frac{\rho U}{B_s} \frac{\partial}{\partial s} \left( r v_\theta \right) + 
r v_\theta\left[ \frac{\partial}{\partial t} \left( \frac{\rho }{ B_s} \right) + \frac{\partial}{\partial s} \left( \frac{\rho U }{ B_s} \right) \right] = \frac{1}{\mu_0}\frac{\partial}{\partial s} \left( r b_\theta \right),    
\end{equation}
and note that the expression inside the square brackets vanishes due to the continuity equation. Application of the solenoidal condition \eqref{solenoidal} to equation \eqref{Eq_B2} leads to the momentum equation \eqref{Eq_Motion}: 
\begin{equation}\label{Eq_B3}
\frac{\partial v_\theta}{\partial t} + B_s U \frac{\partial}{\partial s} \left( \frac{ v_\theta}{B_s} \right) = \frac{B_s^2}{\mu_0 \rho}\frac{\partial}{\partial s} \left( \frac{b_\theta}{B_s} \right).
\end{equation}
\end{appendix}
\end{document}